\documentclass[prl,reprint,showpacs,superscriptaddress,amsmath]{revtex4-1}

\usepackage{bm}
\usepackage{graphicx}
\usepackage{dsfont}
\usepackage{color}

\newcommand{\bra}[1]{\langle #1 | \,}
\newcommand{\ket}[1]{\, | #1 \rangle}

\newcommand{\expv}[1]{\langle #1 \rangle}
\newcommand{\om}{\omega}
\newcommand{\Om}{\Omega}
\newcommand{\ga}{\gamma}
\newcommand{\Ga}{\Gamma}
\newcommand{\de}{\delta}
\newcommand{\De}{\Delta}

\newcommand{\eps}{\epsilon}
\newcommand{\br}{\mathbf{r}}
\newcommand{\mc}[1]{\mathcal{#1}}
\newcommand{\sig}{\hat{\sigma}}
\newcommand{\Sig}{\hat{\Sigma}}
\newcommand{\hOm}{\hat{\Omega}}
\newcommand{\hS}{\hat{S}}
\newcommand{\hs}{\hat{s}}
\newcommand{\halp}{\hat{\alpha}}
\newcommand{\hE}{\hat{\mathcal{E}}}

\begin{document}

\title{Electromagnetically Induced Transparency with Rydberg Atoms}
 
\author{David Petrosyan}
\affiliation{Fachbereich Physik und Forschungszentrum OPTIMAS,
Technische Universit\"at Kaiserslautern, D-67663 Kaiserslautern, Germany}
\affiliation{Institute of Electronic Structure and Laser, 
FORTH, GR-71110 Heraklion, Crete, Greece}

\author{Johannes Otterbach}
\affiliation{Fachbereich Physik und Forschungszentrum OPTIMAS,
Technische Universit\"at Kaiserslautern, D-67663 Kaiserslautern, Germany}
\affiliation{Physics Department, Harvard University,
Cambridge, Massachisets 02138, USA}

\author{Michael Fleischhauer}
\affiliation{Fachbereich Physik und Forschungszentrum OPTIMAS,
Technische Universit\"at Kaiserslautern, D-67663 Kaiserslautern, Germany}


\begin{abstract}
We present a theory of electromagnetically induced transparency
in a cold ensemble of strongly interacting Rydberg atoms. 
Long-range interactions between the atoms constrain the medium 
to behave as a collection of superatoms, each comprising a blockade
volume that can accommodate at most one Rydberg excitation. 
The propagation of a probe field is affected by its two-photon 
correlations within the blockade distance, which are strongly 
damped due to low saturation threshold of the superatoms. 
Our model is computationally very efficient and is in quantitative 
agreement with the results of recent experiment of Pritchard {\it et al.} 
[Phys. Rev. Lett. {\bf 105}, 193603 (2010)]
\end{abstract}


\maketitle

Strong dipole--dipole or van der Waals (VdW) interactions between 
atoms in highly excited Rydberg states \cite{RydAtoms} constitute 
the basis for promising quantum information schemes \cite{rydrev}
and interesting many-body effects 
\cite{Weimer2008,Honer2010,Weimer10,Schachenmayer2010,Pohl2010,Viteau2011}.
Many of these studies utilize the dipole blockade mechanism 
\cite{Lukin2001,Vogt2006,Tong2004,Singer2004,Heidemann2007} 
which suppresses multiple Rydberg excitations 
within a certain interaction (blockade) volume. 
Electromagnetically induced transparency (EIT) \cite{EIT} can translate 
the interactions between Rydberg atoms into sizable interactions between 
single photons \cite{Friedler2005,Petrosyan2008,Gorshkov2011}. 

Recently, several experiments on EIT \cite{adams07,adams08,spreeuw10,adams10},
and the closely related CPT (coherent population trapping) \cite{schempp10}, 
with Rydberg atoms were performed. Strong VdW interactions between the 
atomic Rydberg states were prominently manifest in Ref. \cite{adams10}: 
Increasing the probe field amplitude led to reduction of its 
transmission within the EIT window, which, quite surprisingly, was 
accompanied by negligible broadening and indiscernible shift of the EIT line. 
Here we develop a theoretical model for EIT with Rydberg atoms, whose 
predictions fully reproduce the experimental observations \cite{adams10}.
The crux of our approach is the coarse-grained treatment 
of the atomic medium composed of effective superatoms (SAs), 
with each SA represented by collective states of atoms in 
the blockade volume that can accommodate only one Rydberg 
excitation. A weak probe field propagates through the EIT 
medium with little attenuation, but for a stronger field 
with more than one photon per SA, the excess photons are 
subject to enhanced---essentially two-level atom---absorption.
This leads to the field attenuation with the simultaneous buildup 
of an avoided volume between the probe photons \cite{Gorshkov2011}. 
The inclusion of two-photon correlations is the key feature of our work,
not present in the numerical simulations of \cite{adams10} 
and recent theoretical studies \cite{pohl11} which agreed with  
the experiment at weak probe fields but had significant discrepancies 
for stronger fields. 
Our theory is not limited to weak probe fields and/or low atomic densities, 
yet, despite intrinsic nonlinearity, it is intuitive and numerically 
efficient, amounting to the solution of a pair of coupled differential 
equations for the probe field intensity and its second-order correlation,
in the spirit of the BBGKY hierarchy.

\begin{figure}[b]
\includegraphics[width=8cm]{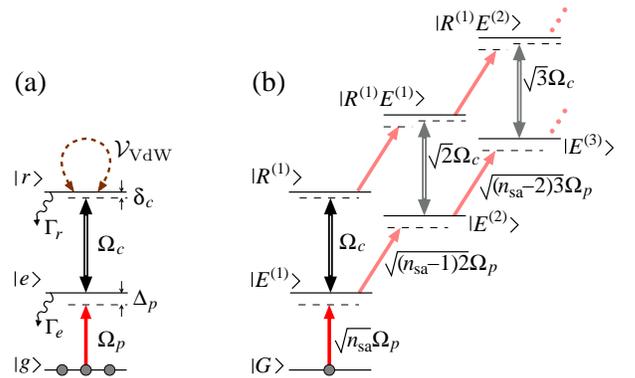}
\caption{
(a) Level scheme of atoms interacting with probe $\Om_{p}$ and 
control $\Om_{c}$ fields with detunings $\De_p$ and $\de_c$. 
$\Ga_e$ and $\Ga_r$ are (population) decay rates of states $\ket{e}$ 
and $\ket{r}$, and $\mc{V}_{\mathrm{VdW}}$ denotes VdW interaction 
between atoms in Rydberg state $\ket{r}$. 
(b) Truncated level scheme of superatom, composed of $n_{\mathrm{sa}}$ atoms, 
with corresponding transition amplitudes due the probe and control fields. 
} 
\label{fig:als}
\end{figure}

Consider an ensemble of $N = \int_V d^3 r \, \rho(\br)$ cold 
atoms of density $\rho(\br)$ in the (quantization) volume $V$
interacting with two optical fields. The quantized probe field $\hE_p$ 
of frequency $\om_p$ acts on the atomic transition between the 
ground $\ket{g}$ and excited $\ket{e}$ states, and the control field 
of frequency $\om_c$ drives the transition $\ket{e} \to \ket{r}$ with 
Rabi frequency $\Om_c$ [see Fig.~\ref{fig:als}(a)]. 
A pair of atoms $i$ and $j$ at positions $\br_i$ and $\br_j$ excited 
to the Rydberg states $\ket{r}$ interact with each other via a VdW 
potential $\hbar \Delta(\br_i -\br_j) = \hbar C_6/|\br_i -\br_j|^6$
\cite{rydcalc}. In the frame rotating with frequencies $\om_{p,c}$, 
the system Hamiltonian 
$\mc{H} = \mc{H}_{\mathrm{a}} + \mc{V}_{\mathrm{af}} + \mc{V}_{\mathrm{VdW}}$ 
contains the unperturbed atomic part, 
$\mc{H}_{\mathrm{a}} = - \hbar \sum_{j}^N [ \De_p \sig_{ee}^j  
+ (\De_p + \de_c) \sig_{rr}^j ]$, and the atom-field and VdW 
interactions, 
$\mc{V}_{\mathrm{af}} =  - \hbar \sum_{j}^N 
[ \hOm_p(\br_j) \sig_{eg}^j + \Om_c \sig_{re}^j + \mathrm{H. c.}]$ and 
$\mc{V}_{\mathrm{VdW}} = \hbar \sum_{i<j}^N 
\sig_{rr}^i \Delta(\br_i -\br_j) \sig_{rr}^j$, 
where $\sig_{\mu \nu}^j \equiv \ket{\mu}_{jj}\bra{\nu}$ is 
the transition operator for atom $j$ at position $\br_j$,
$\De_p = \om_p - \om_{eg}$ and $\de_c = \om_c - \om_{re}$ are
the detunings of the probe and control fields, 
and $\hOm_p = \eta \hE_p$ is the operator of the probe Rabi frequency,
with $\eta = \wp_{ge} \sqrt{\om_p/(2 \hbar \eps_0 V)}$ the atom-field 
coupling strength on the $\ket{g} \to \ket{e}$ transition with dipole
moment $\wp_{ge}$. 

We consider the evolution of the probe field only along its 
propagation $z$ axis, and assume uniform, undepleted control field $\Om_c$.
Using Hamiltonian $\mc{H}$, we obtain Heisenberg-Langevin equations for 
the field $\hE_p(\br)$ and continuous, appropriately averaged, atomic 
$\sig_{\mu \nu}(\br)$ operators:
\begin{subequations}
\label{HLeqs}
\begin{eqnarray}
\left( \partial_t + c \partial_z \right) \hE_p(\br) 
&=& i \eta N \sig_{ge}(\br) , \label{ddzE} \\
\partial_t \sig_{ge}(\br) 
&=& ( i \De_p - \ga_{e}) \sig_{ge}(\br) + i \Om_c^* \sig_{gr}(\br) 
\nonumber \\ & & 
+ i \hOm_p(\br) [\sig_{gg}(\br) - \sig_{ee}(\br)], 
\label{ddte} \\
\partial_t \sig_{gr}(\br) 
&=& [i (\De_2 - \hS(\br)) - \ga_{r} ] \sig_{gr}(\br) 
\nonumber \\ & & 
- i \hOm_p(\br) \sig_{er}(\br) + i \Om_c \sig_{ge}(\br), 
\qquad \label{ddtr}
\end{eqnarray}
\end{subequations}
where $\ga_e \geq \frac{1}{2} \Ga_e$ and $\ga_{r}$ $(\ll \ga_{e})$ are the 
transversal relaxation rates, with the associated noise operators dropped,
$\De_2 = \De_p + \de_c$ is the 
two-photon detuning, and $\hS(\br) \equiv \int  d^3 r' \rho(\br') \, 
\Delta(\br -\br') \sig_{rr}(\br')$ is the total VdW induced shift 
of level $\ket{r}$ for an atom at position $\br$. Since $\hS(\br)$ 
involves integration over all spatial coordinates 
$\br' \in V$, Eqs.~(\ref{HLeqs}) are highly nonlocal. We therefore 
need to contrive an efficient method to evaluate the VdW shift $\hS(\br)$.

We shall be concerned with stationary interaction and drop 
in Eqs.~(\ref{HLeqs}) all the time-derivatives. 
Consider for the moment Eqs.~(\ref{ddte}) and (\ref{ddtr}) 
without the VdW shift $\hS$ and small relaxation $\ga_{r}$ terms.
When $\Om_{p,c} , |\De_{p,2}| < \ga_e$, we can approximate 
the population of Rydberg state $\ket{r}$ by a Lorentzian 
function of $\De_2$: $\expv{\sig_{rr}} \approx 
\expv{\hOm_p^{\dag} \hOm_p} / (|\Om_c|^2 + \De_2^2 \ga_e^2/|\Om_c|^2)$,
with the half-width $w \equiv |\Om_c|^2/\ga_e$. Observe now that an atom 
in Rydberg state $\ket{r}$ would induce VdW shift $\Delta(R)$ of 
level $\ket{r}$ for another atom separated by distance $R$, which 
effectively translates into the two-photon detuning $\De_2$.
The VdW interaction then blocks the excitation of all the atoms 
for which $\Delta(R) \gtrsim w$. This is the essence of the 
Rydberg blockade \cite{Lukin2001}. 
We therefore define the blockade radius $R_{\mathrm{sa}} \simeq \sqrt[6]{C_6/w}$
and call the ensemble of $n_{\mathrm{sa}} = \rho V_{\mathrm{sa}}$ atoms 
within volume $V_{\mathrm{sa}} = \frac{4\pi}{3} R_{\mathrm{sa}}^3$  
``superatom'' (SA). Since in general the atomic density $\rho(\br)$ 
varies with position $\br$, so does $n_{\mathrm{sa}}(\br)$,
but the density of SAs $\rho_{\mathrm{sa}} = V_{\mathrm{sa}}^{-1}
= \frac{3}{4 \pi} \sqrt{|\Om_c|^2/(\ga_e C_6)}$ is constant.    

Each SA can contain only one Rydberg excitation delocalized 
over $V_{\mathrm{sa}}$. We may therefore treat the medium as a collection 
of $N_{\mathrm{sa}} = \rho_{\mathrm{sa}} V$ SAs at positions $\br_j$, 
which implies a spatial coarse-graining with the grain size $2 R_{\mathrm{sa}}$
\cite{supatom,RydCorrels}.
The total VdW shift $\hS(\br)$ at position $\br$ can then be expressed as
\begin{equation}
\hS(\br) \approx \sum_j^{N_{\mathrm{sa}}} \De(\br - \br_j ) 
\Sig_{RR}(\br_j) = \bar{\De} \Sig_{RR}(\br) + \hs(\br) , \label{SRr}
\end{equation}
where $\Sig_{RR}(\br_j)$ is the projector onto the Rydberg excitation
of SA at $\br_j$. 
The physical meaning of the first term on the rhs of Eq.~(\ref{SRr}) 
is that an excited SA at $\br_j \simeq \br$ [$\Sig_{RR}(\br) \to 1$] 
induces divergent VdW shift averaged over the SA volume: $\bar{\De} \simeq 
\frac{1}{V_{\mathrm{sa}}} \int_{V_{\mathrm{sa}}} \De (\br') d^3 r' \to \infty$. 
Actually, for a small cut-off in the interatomic separation, 
$|\bar{\De}| \gg \ga_e$ is finite but very large, which is the 
only relevant property. The last term $\hs(\br) \equiv 
\sum_{j\neq j_{\br}}^{N_{\mathrm{sa}}} \De(\br - \br_j) \Sig_{RR}(\br_j)$
describes the VdW shift induced by the external SAs outside the volume 
$V_{\mathrm{sa}}^{(\br)}$ centered at $\br$. It can be evaluated by 
replacing the summation by an integration over the entire volume $V$, 
excluding the SA at $\br$, which, upon using the mean-field approximation,
yields a small shift $\expv{\hs(\br)} = \frac{w}{8} \expv{\Sig_{RR}(\br)}$.

Assuming that the probe field $\hOm_{p}$ varies little over distance
$\sim R_{\mathrm{sa}}$, we can describe the dynamics of individual 
SAs in terms of collective states and operators defined within 
the blockade volume $V_{\mathrm{sa}}$. 
The level scheme of SA is shown in Fig.~\ref{fig:als}(b): 
$\ket{G} = \ket{g_1,g_2, \ldots,g_{n_{\mathrm{sa}}}}$ is the ground state, 
and $\ket{R^{(1)}} = \frac{1}{\sqrt{n_{\mathrm{sa}}}}
\sum_j^{n_{\mathrm{sa}}} \ket{g_1,g_2, \ldots,r_j,\ldots,g_{n_{\mathrm{sa}}}}$
is the single collective Rydberg excitation state, while $\ket{E^{(k)}}$ 
are the properly symmetrized (Dicke) states with $k$ atoms in $\ket{e}$. 
The corresponding transition amplitudes 
$\bra{E^{(1)}} \mc{V}_{\mathrm{af}} \ket{G} = \sqrt{n_{\mathrm{sa}}} \hOm_p$,  
$\bra{R^{(1)}} \mc{V}_{\mathrm{af}} \ket{E^{(1)}} = \Om_c$, etc.,
depend of the number of atoms $n_{\mathrm{sa}}$ in $V_{\mathrm{sa}}$.
In order to calculate $\Sig_{RR}$, we now proceed along 
the lines similar to the single atom treatment. 
Starting with the SA in $\ket{G}$, we adiabatically eliminate all 
the excited states $\ket{E^{(k)}}$ having large widths $\sim k \gamma_e$.  
Note that state $\ket{R^{(1)}}$ is reached from $\ket{G}$ by 2-photon 
transition, while all the other states $\ket{R^{(1)} E^{(k)}}$ require 
$2+k$ photon transitions; therefore their adiabatic elimination 
affects little the populations of $\ket{G}$ and $\ket{R^{(1)}}$. 
We then obtain for the SA operators 
$\Sig_{GR} \equiv \ket{G} \bra{R^{(1)}} 
= \Om_c \sqrt{n_{\mathrm{sa}}} \hOm_p \Sig_{GG}/
[(\De_p + i \ga_e) \De_2 - |\Om_c|^2]$ and $\Sig_{RR} = \Sig_{RG} \Sig_{GR}$.
To account for possible saturation of transition 
$\ket{G} \to \ket{R^{(1)}}$ when the number density of 
probe photons $\rho_{\mathrm{phot}}$ is comparable to, or larger than, 
the density of SAs $\rho_{\mathrm{sa}}$, we take 
$\Sig_{GG} + \Sig_{RR} = \mathds{1}$, which finally 
yields \cite{SigmaRRfootnote}
\begin{equation}
\Sig_{RR} = \frac{|\Om_c|^2 n_{\mathrm{sa}} \hOm_p^{\dag} \hOm_p}
{|\Om_c|^2 n_{\mathrm{sa}} \hOm_p^{\dag} \hOm_p +
[|\Om_c|^2 - \De_p \De_2]^2 + \De_2^2 \ga_e^2} . \label{SigRR}
\end{equation}

We next examine the probe field propagation in the atomic medium. 
For moderate Rabi frequency $\Om_{p} < \ga_e$ and number density 
of photons $\rho_{\mathrm{phot}} \ll \rho$, we can assume linear 
response of individual atoms to the applied field, setting 
$\sig_{ee},\sig_{er} \to 0$ and $\sig_{gg} = \mathds{1}$. We then 
arrive at the propagation equation for the probe field amplitude, 
$\partial_z \hE_p = i \frac{\kappa}{2} \halp \hE_p$, where 
$\kappa = \varsigma_0 \rho$ is the resonant (intensity) absorption 
coefficient proportional to the atomic absorption cross-section 
$\varsigma_0 = \om_p |\wp_{ge}|^2/(\hbar \eps_0 c \ga_{e})$, while 
\begin{eqnarray}
\halp(\br) &=& 
\Sig_{RR}(\br) \frac{i \ga_{e}} {\ga_{e} - i \De_p} 
+ [ 1 - \Sig_{RR}(\br)] 
\nonumber \\ & & \times
\frac{ i\ga_{e}}{\ga_{e} - i \De_p + |\Om_c|^2 
[\ga_{r} - i (\De_2 - \expv{\hs(\br)})]^{-1} } \label{alphap}
\end{eqnarray}
is the operator-valued polarizability. 
Here the first fraction is the polarizability $\alpha_{\mathrm{TLA}}$ 
of a two-level atom, while the second fraction, barring the small 
shift $\expv{\hs(\br)}$, is the usual EIT polarizability 
$\alpha_{\mathrm{EIT}}$ \cite{EIT}. Physically, if the SA at position 
$\br$ contains a Rydberg excitation [$\Sig_{RR}(\br) \to 1$], 
the two-photon detuning is shifted out of the EIT transparency window 
[$\bar{\De} \gg \ga_e$] and the probe field $\hE_p$ sees an absorbing 
two-level system; if no Rydberg excitation is present, the medium 
response is that of usual EIT with a small mean-field shift due 
to the VdW interaction with the external SAs.
Then the expectation value of the probe field intensity obeys the equation
\begin{equation}
\partial_z \expv{\hE_p^{\dag}(\br) \hE_p(\br)} = - \kappa(\br) 
\expv{\hE_p^{\dag}(\br) \mathrm{Im}[\halp(\br)] \hE_p(\br)} . \label{Ipz}
\end{equation}
Note that factorizing out $\mathrm{Im}[\expv{\halp(\br)}]$
in a mean-field sense would amount to neglecting the essential 
two-particle quantum correlations \cite{Gorshkov2011} originating 
from nonlinear response of the atoms to the Rydberg excitations.
We therefore proceed more carefully and replace $\halp(\br)$ in 
Eq.~(\ref{Ipz}) by its expectation value {\em conditioned} upon 
the presence of photon at $\br$, denoted by $\expv{\cdot}_{\br}$, 
\begin{equation}
\expv{\halp(\br)}_{\br} = 
\expv{ \Sig_{RR}(\br)}_{\br} \alpha_{\mathrm{TLA}} 
+ [ 1 - \expv{ \Sig_{RR}(\br)}_{\br}] \alpha_{\mathrm{EIT}} . \label{alphaS}
\end{equation}
The conditional Rydberg population $\expv{\Sig_{RR}}_{\br}$ of the SA 
at $\br$ is obtained from Eq.~(\ref{SigRR}) by the replacement
$\hOm_p^{\dag}(\br) \hOm_p(\br) 
\to \expv{\hOm_p^{\dag}(\br) \hOm_p(\br)} \, g_p^{(2)}(\br)$, where
the probe field intensity correlation function 
$g_p^{(2)}(\br) = \frac{\expv{\hE_p^{\dag}(\br) \hE_p^{\dag}(\br)
\hE_p(\br)\hE_p(\br)}}{\expv{\hE_p^{\dag}(\br) \hE_p(\br)} 
\expv{\hE_p^{\dag}(\br) \hE_p(\br)}}$
quantifies the probability of having simultaneously at least 
two photons in the blockade volume $V_{\mathrm{sa}}^{(\br)}$.
The field intensity is now coupled to its two-photon correlation
$g_p^{(2)}(\br)$ which in turn evolves upon propagation. 
Note that linear, e.g. bare EIT, response of the medium does not 
change the correlation function of the propagating field,
and only nonlinear, i.e. conditional, absorption 
$\propto \mathrm{Im} [\expv{\alpha(\br)} - \alpha_{\mathrm{EIT}}]$
modifies $g_p^{(2)}$, which therefore obeys the equation of motion 
\cite{gp2drvfootnote}
\begin{equation}
\partial_z g_p^{(2)}(\br) 
= - \kappa(\br) \expv{ \Sig_{RR}(\br)} 
\mathrm{Im} [\alpha_{\mathrm{TLA}} - \alpha_{\mathrm{EIT}}] g_p^{(2)}(\br) . 
\label{g2}
\end{equation}
Hence, within the EIT transparency window, where
$\mathrm{Im} [\alpha_{\mathrm{EIT}}] \simeq 0$, the correlations 
between the photon pairs with relative distance smaller than the 
blockade (SA) radius decay with the rate proportional to the 
probability of SA excitation $\expv{ \Sig_{RR}}$ and the absorption 
rate of a two-level system $\mathrm{Im} [\alpha_{\mathrm{TLA}}]$.
We note that our treatment involves only single transverse mode 
of the probe field, which is effectively defined by the SA cross-section. 
If, however, during propagation there is strong mixing of the 
transverse modes, it would preclude the buildup of (anti)correlations
between the photons.

Given the input field ``intensity'' $I_p \equiv \expv{\hOm_p^{\dag}\hOm_p}$ 
and its correlation function $g_p^{(2)}$ 
[for ``classical'' coherent field $g_p^{(2)}(0)= 1$],
we then use the following stochastic procedure to spatially integrate 
the coupled coarse-grained Eqs.~(\ref{Ipz})-(\ref{g2}) for $z \in [0,L]$: 
We divide the propagation distance $L$ into $L/(2 R_{\mathrm{sa}})$ 
intervals corresponding to SAs,  
and for $z$ within each SA we determine via Monte-Carlo
sampling of $\expv{\Sig_{RR}(\br)}_{\br}$ whether the SA is excited, 
$\Sig_{RR}(\br) \to 1$, or not, $\Sig_{RR}(\br) \to 0$. 
We then average over several independent realizations. 
The limit of infinitely many such realizations 
corresponds to continuous polarizability of Eq.~(\ref{alphaS}).

We employ our theory to simulate the experiment of 
Ref.~\cite{adams10} with an ensemble of cold $^{87}$Rb atoms: 
$\ket{g} \equiv 5 S_{1/2} \ket{F=2,m_F=2}$,
$\ket{e} \equiv 5 P_{3/2} \ket{F=3,m_F=3}$ with 
$\Ga_e = 3.8 \times 10^{7} \:$s$^{-1}$, and $\ket{r} \equiv 60 S_{1/2}$ 
with $\Ga_r = 5 \times 10^{3} \:$s$^{-1}$ and 
$C_6/2 \pi = 1.4 \times 10^{11} \:$s$^{-1}\mu$m$^6$ 
\cite{rydcalc} corresponding to repulsive VdW interactions. 
$\ga_{e,r}$ also include the one- and two-photon laser linewidths 
$\de \om_{1,2}/2 \pi \simeq (5.7,11) \times 10^4\:$s$^{-1}$.
The atomic density is $\rho(z) = \rho_0 \exp[-(z-z_0)^2/2 \sigma_{\rho}^2]$
with peak $\rho_0 = 1.32 \times 10^7\:$mm$^{-3}$ and half-width
$\sigma_{\rho} = 0.7\:$mm; indistinguishable results are obtained 
for homogeneous ensemble of density 
$\bar{\rho} = 1.2 \times 10^7\:$mm$^{-3}$ and length $L=1.3\:$mm, 
leading to the resonant optical depth of $\bar{\kappa} L = 4.524$. 
The control field $\Om_c/2 \pi = 2.25 \times 10^6\:$s$^{-1}$ \cite{Rabifootnote}
is slightly detuned by $\de_c/2 \pi = -10^5\:$s$^{-1}$. 
The corresponding blockade radius is $R_{\mathrm{sa}} \simeq 6.6\:\mu$m 
and each SA contains on average $\bar{n}_{\mathrm{sa}} \simeq 14.7$ 
atoms. We emphasize that our simulations are insensitive to 
moderate variations ($\pm 20\%$) of the SA volume $V_{\mathrm{sa}}$ 
and the number of atoms $n_{\mathrm{sa}}$ ($\simeq 14 \pm 3$) it contains.

\begin{figure}[t]
\includegraphics[width=6.4cm]{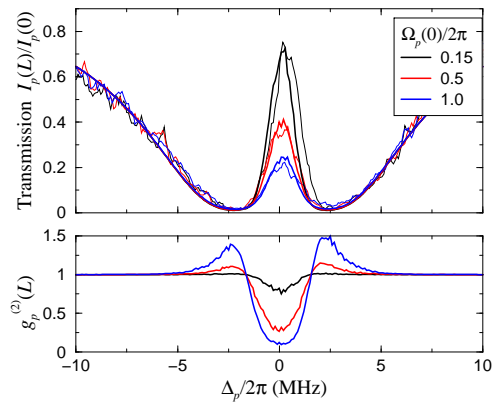}
\caption{
Top: Probe field transmission $I_p(L)/I_p(0)$ versus 
detuning $\De_p$, for different input intensities corresponding 
to $\Om_p(0)/2\pi = 0.15,0.5,1.0\:$MHz.
Thin lines are experimental curves from \cite{adams10},  
thicker lines are obtained via stochastic simulations of 
Eqs.~(\ref{Ipz})-(\ref{g2}) averaged over 10 independent realizations. 
Bottom: The corresponding intensity correlation functions $g_p^{(2)}(L)$. 
} 
\label{fig:ran}
\end{figure}

In Fig.~\ref{fig:ran} we compare the transmission spectra for different 
input probe intensities with the corresponding plots of Ref.~\cite{adams10}. 
Already for $\Om_p(0)/2\pi \gtrsim 0.1\:$MHz the VdW interaction induced 
nonlinearities play an important role. The agreement between our stochastic 
simulations and the experiment is remarkable. We also show the local 
intensity correlation $g_p^{(2)}(L)$ at the exit from the medium.

\begin{figure}[t]
\includegraphics[width=8.0cm]{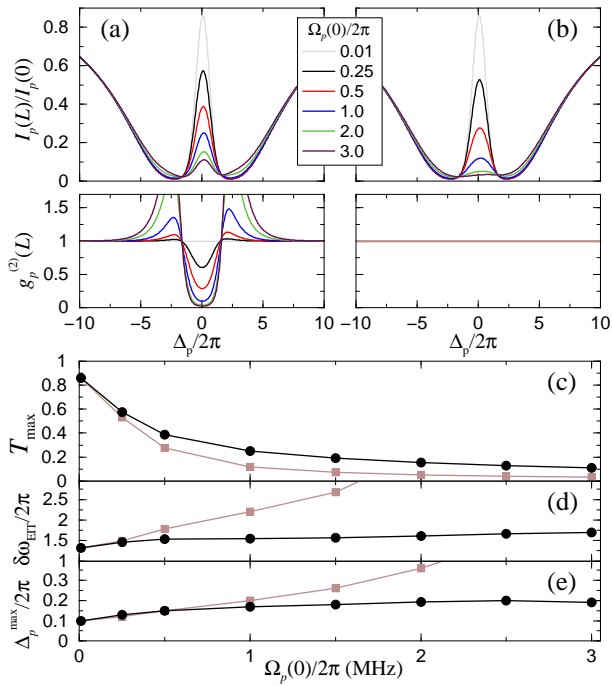}
\caption{
(a) Transmission (top) and intensity correlation (bottom) spectra of 
the probe field, for various input intensities. 
(b) The same, but setting $g_p^{(2)}(z) = 1 \, \forall \, z \in[0,L]$.
(c) Peak probe transmission $T_{\textrm{max}}$ around the EIT line center,
(d) EIT linewidth $\de \om_{\mathrm{EIT}}$ (FWHM), and 
(e) detuning $\De_p^{\textrm{max}}$ at the maximum $T_{\textrm{max}}$, 
versus the input probe Rabi frequency $\Om_p(0)$. 
The black lines (circles) correspond to case (a) 
and the brown lines (squares) to (b). 
$\De_p, \de \om_{\mathrm{EIT}}, \Om_p$ are in MHz. } 
\label{fig:ravcomb}
\end{figure}

Figure~\ref{fig:ravcomb} summarizes the results of our simulations 
involving the continuous polarizability of Eq.~(\ref{alphaS}). 
The weak field of $\Om_p/2\pi \lesssim 0.01\:$MHz encounters
linear EIT response. Increasing the input probe intensity leads
to lesser transmission through the EIT window ($\De_2 \sim 0$) 
and to small mean-field shift and broadening of the EIT line. 
This is due to the higher probability of two or more 
photons, exciting Rydberg states $\ket{r}$, to be at the same SA.
The induced large VdW level shift $\bar{\De}$ results in strong 
photon absorption, simultaneously reducing the photon coincidence 
probability within the SA volume $V_{\mathrm{sa}}$.
Hence, both $I_p(z)$ and $g_p^{(2)}(z)$ decay, but once $g_p^{(2)}(z) \ll 1$, 
the attenuation of the probe field intensity $I_p(z)$ slows down. 
Eventually $I_p = \expv{\hOm_p^{\dag} \hOm_p}$ saturates at a value 
corresponding to less than one photon per SA, 
$\rho_{\mathrm{phot}} \lesssim \rho_{\mathrm{sa}}$, 
with vanishing coincidence probability. 
With $\rho_{\mathrm{phot}} = \hbar \eps_0 c I_p/(2 \wp_{ge}^2 \om_p v)$, 
where $v = 2|\Om_c|^2/(\bar{\kappa} \ga_e)$ ($\simeq 6000\:$m/s) 
is the probe group velocity within the EIT window  
$|\De_2| \lesssim \de \om_{\mathrm{EIT}}$, we have that
$\rho_{\mathrm{phot}} = (\rho/4) \expv{\hOm_p^{\dag} \hOm_p} /|\Om_c|^2$
and the maximal saturation intensity is 
$I_p^{\mathrm{max}} \simeq (4 \rho_{\mathrm{sa}}/ \rho) |\Om_c|^2$.
In the medium the photons are anticorrelated (antibunched) within the 
temporal window of $\de t \simeq 2 R_{\mathrm{sa}}/v$ ($\simeq 1.6\:$ns), 
which does not change when they leave the medium for free space.

Had we not taken into account the probe field intensity correlation, 
equivalent to setting $g_p^{(2)}(z) = 1 \, \forall \, z \in[0,L]$, 
Fig.~\ref{fig:ravcomb}(b), we would have faster, exponential decay 
of $I_p(z)$, unrestrained by the buildup of avoided volume between 
the photons, as well as sizable shift and broadening of the EIT line,
which contradict the observations of \cite{adams10}. 
  
Outside the EIT window, around the Autler-Townes doublet 
$\De_2 \sim \pm \Om_c$, the probe is strongly absorbed, 
$\mathrm{Im} [\expv{\alpha}] \simeq 1$, but 
the correlation function is amplified, since in Eq.~(\ref{g2})
$\mathrm{Im} [\alpha_{\mathrm{TLA}} - \alpha_{\mathrm{EIT}}] < 0$.
In other words, linear absorption is larger than the conditional
absorption, which results in photon bunching but very low flux. 

We finally note that for relatively strong input fields 
$\Omega_p(0) \lesssim \ga_e$ of Fig.~\ref{fig:ravcomb}, 
the validity of linear response of the atoms inherent 
in polarizability of Eq.~(\ref{alphap}) may not be 
\textit{a priori} justified. 
This is indeed the case for an optically thin atomic medium. 
But in the optically thick medium, within a few absorption 
lengths, even a strong probe field and its two-photon correlation
function quickly decay to the level at which the above approximation
is justified. 

To conclude, EIT via atomic Rydberg states is suppressed by collective 
Rydberg excitations of SAs which depend on the local probe field intensity 
and its two-particle correlation within the SA (blockade) volume. 
For strong input fields, the buildup of anticorrelations between 
the photons upon propagation through the medium leads to 
the saturation of transmitted field intensity to a value 
corresponding to one photon per blockade volume. 
Conversely, suitably antibunched input fields 
should exhibit large transmission affected only by small linear absorption.

In a one-dimensional configuration, the spatial correlations between 
the photons in the medium translate at the output into temporal 
correlations in free space, which can be measured by coincidence detection.
The limit of maximal saturation intensity $I_p^{\mathrm{max}}$ of the 
transmitted through the EIT window field then corresponds to a train 
of non-overlapping single-photon pulses with the temporal separation 
$\de t$ of a few ns.

\begin{acknowledgments}
We are grateful to J.D. Pritchard and C.S. Adams 
for sharing with us the experimental details. 
We thank A. Gorshkov, Th. Pohl and M.D. Lukin for stimulating discussions. 
This work was supported by the Humboldt Foundation (D.P.),
the Harvard Quantum Optics Center (J.O.), and SFB TR49 (M.F.).
\end{acknowledgments}


\begin{thebibliography}{99}
\bibitem{SigmaRRfootnote}
Equation~(\ref{SigRR}) can be derived without resorting to symmetric
states and operators of SA, which is however more involved.
\bibitem{Rabifootnote}
Our definition of the Rabi frequencies $\Om_{p,c}$ differ from that in 
\cite{adams10} by a factor of $\frac{1}{2}$.
\end{thebibliography}
\end{document}